\documentclass[11pt,twoside,epsf]{article}
\usepackage[dvips]{graphicx}

\begin{document}

\setcounter{figure}{0}
\setcounter{section}{0}
\setcounter{equation}{0}
\setcounter{table}{0}

{ \huge \bf Covariant Mixmaster Dynamics }

\vspace{1.cm}
{ \Large Giovanni Imponente and Giovanni Montani
\vspace{.1cm}}

{\it \Large ICRA-International Center for Relativistic Astrophysics \\ 
and Physics Department, G9, University of Rome ``La Sapienza'', piazza A.Moro 5 00185 Rome, Italy \\
 }{\rm e-mail: imponente@icra.it}

\begin{abstract}
We provide a Hamiltonian analysis of the Mixmaster Universe dynamics 
on the base of a standard Arnowitt-Deser-Misner Hamiltonian approach,
showing the covariant nature of its chaotic behaviour with respect to the 
choice of any time variable, from the point of view either of the dynamical systems theory, 
either of the statistical mechanics one.

\end{abstract}

\section{Introduction}
Since Belinski-Kalatnikov-Lifshitz (BKL) derived the oscillatory regime which characterizes 
 the behaviour of the Bianchi type VIII and IX cosmological models \cite{BKL70} 
 (the so-called Mixmaster universe \cite{M69}) near a physical singularity, a wide literature 
faced over the years this subject in order to provide the best possible understanding of 
the resulting chaotic dynamics. 

The research activity developed overall in two different, but related, directions: on one 
hand the dynamical analysis was devoted to remove the limits of the BKL approach due
 to its discrete nature, on the other one to get a better characterization of 
the Mixmaster chaos (especially in view of its properties of covariance). \\
The first line of investigation provided satisfactory representations of the Mixmaster 
dynamics in terms of continuous variables 
(leading to the construction of an invariant measure for the system \cite{CB83}, \cite{KM97}). \\
The efforts in the second direction found non-trivial difficulties due to the impossibility, 
to apply the standard chaos indicators to relativistic systems. 

The existence of difficulties related to the covariance of various approaches 
prevented, up to now, to say a definitive word 
about the Mixmaster chaoticity (apart from the indication provided by \cite{CL97}, see also \cite{ML00}). 

The aim of this work is to show how in the points of view of the theory of 
dynamical systems and statistical mechanics
the representation of the Mixmaster chaoticity is independent on the choice of a time gauge. 

The description of the system evolution as a ``stochastic scattering'' isomorphic to a billiard 
on the Lobachevsky space can be constructed independently of the choice of a time variable, 
simply providing very general Misner-Chitr\'e-like coordinates. 

On the other hand, we show how the derivation of an invariant measure for the Mixmaster 
model (performed in \cite{KM97,M00} 
within the framework of the statistical mechanics) can be extended to a generic 
time gauge. In fact, asymptotically close to the cosmological 
singularity, the Mixmaster dynamics can be modeled by a two-dimensional 
point-Universe randomizing in a closed domain with fixed ``energy'' 
(just the ADM kinetic energy); since it is natural to represent such a 
system by a microcanonical ensemble, then the corresponding invariant 
measure is induced by the Liouville one.

Up to the limit of the adopted approximation on the form of the potential term, our analysis shows, without any ambiguity, that the Mixmaster stochasticity can not be removed by any redefinition of the time variable.

\section{The Hamiltonian Formulation}\label{hamiltonian}

The geometrical structure of the Bianchi type VIII and IX spacetimes, i.e. of the so-called Mixmaster Universe models, is summarized by the line element \cite{BKL70}

\begin{equation} 
ds^2=-{N(\eta)}^2d{\eta}^2+e^{2\alpha}\left(e^{2\beta}\right)_{ij}\sigma^i \sigma^j 
\label{a} 
\end{equation} 
where $N(\eta)$ denotes the lapse function, $\sigma^i$ are the dual 1-forms associated with the anholonomic basis \footnote{ The dual 1-forms of the considered models are given by: \\[1em]

(Bianchi ~VIII): $\left\{ 
\begin{array}{lll} 
\sigma ^1 =-\sinh \psi \sinh\theta d\phi~ + ~\cosh \psi d\theta \\
\sigma ^2= -\cosh \psi \sinh\theta d\phi ~+~\sinh \psi d\theta \\
\sigma ^3=~\cosh\theta d\phi ~+ ~d \psi 
\end{array} \right.$ 
\\[1em]

(Bianchi ~IX): \quad  $\left\{ 
\begin{array}{lll} 
\sigma ^1 = ~\sin \psi \sin\theta d\phi ~+~\cos \psi d\theta \\
\sigma ^2 = -\cos \psi \sin\theta d\phi ~+~\sin \psi d\theta \\
\sigma ^3 = ~\cos\theta d\phi ~+~d \psi  
\end{array} \right.$
}
and $\beta_{ij}$ is a traceless $3\times3$ symmetric matrix  ${\rm diag}(\beta_{11},\beta_{22},\beta_{33})$; $\alpha$, $N$, $\beta_{ij}$ are functions of $\eta$ only. Parameterizing the matrix $\beta_{ij}$ by the usual Misner variables \cite{M69}
\begin{equation}
\beta_{11}=\beta_+ + \sqrt3 \beta_- \, , \quad \beta_{22}=\beta_+ - \sqrt3 \beta_-  \, , \quad \beta_{33}=-2 \beta_+  
\label{a1}
\end{equation} 
the dynamics of the Mixmaster model is described by a canonical variational principle
$
\delta I=\delta\int L\, d\eta=0 ,
$
with Lagrangian $L$ 

\begin{equation} 
L=\frac{6 D}{N}\left[{-{\alpha}^{\prime}}^2+{{\beta_+}^{\prime}}^2+{{\beta_-}^{\prime}}^2\right]- \frac{N}{D}V\left(\alpha, \beta_+ , \beta_-\right).
\label{r} 
\end{equation}  
Here ${()}^{\prime} = \frac{d}{d\eta}$, $D\equiv \det e^{\alpha +\beta_{ij}}=e^{3\alpha}$  and the potential $V\left(\alpha, \beta_+ , \beta_-\right)$ reads

\begin{equation} 
V=\frac{1}{2} \left( D^{4H_1}+D^{4H_2}+D^{4H_3}\right) -D^{2H_1 +2H_2}\pm D^{2H_2 +2H_3}\pm D^{2H_3 +2H_1} ,
\label{a2} 
\end{equation}  
where $(+)$ and $(-)$ refers respectively to Bianchi type VIII and IX, and the anisotropy parameters $H_i ~(i=1,2,3)$ denote the functions \cite{KM97}
\begin{equation} 
H_1 = \frac{1}{3}+ \frac{\beta_+ + \sqrt{3} \beta_-}{3 \alpha} \, , \quad H_2 = \frac{1}{3}+ \frac{\beta_+ - \sqrt{3} \beta_-}{3 \alpha} \, ,\quad H_3 =\frac{1}{3}- \frac{2\beta_+}{3 \alpha}  \, .
\label{ssaa}
\end{equation} 
In the limit $D\rightarrow 0$ the second three terms of the above potential turn out to be negligible with respect to the first one.
Let's introduce the new (Misner-Chitr\'e-like) variables 
\begin{equation}
\alpha = -e^{f\left(\tau\right)}\xi \, , \quad \beta_+ = e^{f\left(\tau\right)}\sqrt{\xi^2 -1}\cos \theta \, , \quad \beta_- =e^{f\left(\tau\right)}\sqrt{\xi^2 -1}\sin \theta  \, , 
\label{f2}
\end{equation} 
with $f$ denoting a generic functional form of $\tau$, $1\le \xi <\infty$ and $0\le \theta < 2 \pi$. Then the Lagrangian (\ref{r}) reads
\begin{equation} 
L=\frac{6 D}{N} \left[ \frac{{\left(e^f {\xi}^{\prime}\right)}^2}{\xi ^2 -1} +{\left(e^f {\theta}^{\prime}\right)}^2\left(\xi ^2 -1\right) -{{\left(e^f\right)}^{\prime}}^2 \right]  -\frac{N}{D}V \left( f\left(\tau\right), \xi, \theta \right) .
\label{g2} 
\end{equation} 
In terms of $f\left(\tau\right)$, $\xi$ and $\theta$ we have
\begin{equation} 
D= exp\left\{ -3 \xi \cdot e^{f\left(\tau\right)} \right\}
\label{h} 
\end{equation} 
and since $D \rightarrow 0$ toward the singularity, independently of its particular form, in this limit $f$ must approach infinity.
The Lagrangian (\ref{r}) leads to the conjugate momenta 
\begin{equation}
p_{\tau}=-\frac{12 D}{N}{\left(e^f \cdot \frac{df}{d\tau}\right)}^2  {\tau}^{\prime}   \, , \quad 
p_{\xi} = \frac{12 D}{N}e^{2f} \frac{{\xi}^{\prime}}{{\xi}^2 -1} \, , \quad 
p_{\theta}= \frac{12 D}{N}e^{2f} {\theta}^{\prime}\left({\xi}^2 -1\right)   
\label{i}
\end{equation} 
which by a Legendre transformation make the initial variational principle assume the Hamiltonian form  
\begin{equation} 
\delta \int \left(   p_{\xi} {\xi}^{\prime} +  p_{\theta} {\theta}^{\prime}+    p_{\tau}  {\tau}^{\prime} - \frac{Ne^{-2f}}{24 D} {\cal H}                            \right) d\eta =0 ,
\label{m} 
\end{equation} 
being 
\begin{equation} 
{\cal H} = -\frac{{p_{\tau}}^2}{\left(\frac{df}{d\tau}\right)^2} + 
{p_{\xi}}^2\left(\xi ^2 -1\right) +\frac{{p_{\theta}}^2}{\xi ^2 -1} +24 V e^{2f}  .
\label{n} 
\end{equation}

By variating (\ref{m}) with respect to $N$ we get the constraint ${\cal H} =0$, which solved provides 
\begin{equation} 
\label{n2}
-p_{\tau}\equiv \frac{df}{d\tau}\cdot {\cal H}_{ADM} = \frac{df}{d\tau} \cdot \sqrt{\varepsilon ^2 +24 V e^{2f}}
\end{equation}
where
\begin{equation}
\varepsilon ^2 = \left({\xi}^2 -1\right){p_{\xi}}^2 +\frac{{p_{\theta}}^2}{{\xi}^2 -1} \, .
\label{d2} 
\end{equation} 
In terms of (\ref{n2}) the variational principle (\ref{m}) reduces to 
\begin{equation} 
\delta \int \left(   p_{\xi} {\xi}^{\prime} +  p_{\theta} {\theta}^{\prime} - {f}^{\prime}{\cal H}_{ADM} \right) d\eta =0\, .
\label{q} 
\end{equation} 
Since the equation for the temporal gauge actually reads
\begin{equation} 
N\left(\eta\right)= \frac{12 D}{{\cal H}_{ADM}} e^{2f} \frac{df}{d\tau} {\tau}^{\prime} \, ,
\label{rs} 
\end{equation} 
our analysis remains fully independent of the choice of the time variable until the form of $f$ and ${\tau}^{\prime}$ is not fixed.

The variational principle (\ref{q}) provides  the Hamiltonian equations for ${\xi}^{\prime}$ and ${\theta}^{\prime}$ 
\footnote{In this study the corresponding equations for $p^{\prime}_{\xi}$ and $p^{\prime}_{\theta}$ are not relevant.}
\begin{equation}
\label{s}
{\xi}^{\prime}= \frac{f^{\prime}}{{\cal H}_{ADM}}\left(\xi ^2 -1\right)p_{\xi}  \, , \qquad        
{\theta}^{\prime}= \frac{f^{\prime}}{{\cal H}_{ADM}} \frac{p_{\theta}}{\left(\xi ^2 -1\right)} \, .
\end{equation}
Furthermore can be straightforward derived the important relation
\begin{equation} 
\frac{d\left({\cal H}_{ADM}f^{\prime}\right)}{d\eta} = \frac{\partial \left({\cal H}_{ADM}f^{\prime}\right)}{\partial\eta} \Longrightarrow \frac{d\left({\cal H}_{ADM}f^{\prime}\right)}{df} = \frac{\partial \left({\cal H}_{ADM}f^{\prime}\right)}{\partial f} \, ,
\label{t} 
\end{equation} 
i.e. explicitly
\begin{equation} 
\frac{\partial{\cal H}_{ADM}}{\partial f}=\frac{e^{2f}}{2 {\cal H}_{ADM}} 24\cdot \left( 2V+ \frac{\partial V}{\partial f} \right)\, .
\label{u} 
\end{equation} 
In this reduced Hamiltonian formulation, the functional $f\left(\eta\right)$ plays simply the role of a parametric function of time and actually the anisotropy parameters $H_i$ $(i=1,2,3)$ are functions of the variables $\xi, \theta$ only 

\begin{eqnarray} 
\label{v4}
H_1 &=& \frac{1}{3} - \frac{\sqrt{\xi ^2 - 1}}{3\xi }\left(\cos\theta + \sqrt{3}\sin\theta \right)   \nonumber \\
H_2 &=& \frac{1}{3} - \frac{\sqrt{\xi ^2 - 1}}{3\xi }\left(\cos\theta - \sqrt{3}\sin\theta \right)  \\
H_3 &=& \frac{1}{3} + 2\frac{\sqrt{\xi ^2 - 1}}{3\xi } \cos\theta \, . \nonumber
\end{eqnarray}

Finally, toward the singularity ($D \rightarrow 0$ i.e. $f \rightarrow \infty$) by the expressions (\ref{a2}, \ref{h}, \ref{v4}), we see that \footnote{By $O()$ we mean terms of the same order of the inclosed ones.}
\begin{equation}
\frac{\partial V}{\partial f} = O\left(e^f V\right) \, .
\label{z0}
\end{equation} 
Since in the domain $\Gamma_H$ all the $H_i$ are simultaneously greater than 0, the potential term $U\equiv e^{2f} V$ can be modeled by the potential walls

\begin{eqnarray}
\label{aa}
U_\infty = 
&\Theta _\infty \left(H_l\left(\xi, \theta\right)\right) + \Theta _\infty \left(H_m\left(\xi, \theta\right)\right) + \Theta _\infty \left(H_n\left(\xi, \theta\right)\right) \, \\ 
 &\Theta _\infty \left(x\right) = \biggl\{ 
\begin{array}{lll} 
+ \infty & if & x < 0 \\ 
\quad 0 & if & x >  0 
\end{array} \nonumber
\end{eqnarray}
therefore in $\Gamma_H$ the ADM Hamiltonian becomes (asymptotically) an integral of motion
\begin{equation}
\forall \{\xi, \theta\}\in{ \Gamma_H} \quad
\left\{ 
\begin{array}{lll} 
{\cal H}_{ADM}= \sqrt{\varepsilon ^2 +24\cdot U} \cong \varepsilon =E =const. \\ 
\frac{\partial {\cal H}_{ADM}}{\partial f} =\frac{\partial E}{\partial f} =  0  \, .
\end{array}
\right .
\label{bb}
\end{equation}

The key point for the use of the Misner-Chitr\'e-like variables relies on the independence of the time variable for the anisotropy parameters $H_i$.

\section{The Jacobi Metric and the Billiard Representation}

Since above we have shown that asymptotically to the singularity ($f~\rightarrow~\infty$, i.e. $\alpha\rightarrow-\infty$) $d{\cal H}_{ADM}/df=0$ i.e. ${\cal H}_{ADM} =\epsilon =E=const.$, the variational principle (\ref{q}) reduces to
\begin{equation}
\delta \int \left( p_{\xi} d\xi + p_{\theta} d\theta -Edf \right) =\delta \int \left(  p_{\xi} d\xi + p_{\theta} d\theta \right)=0 \quad ,
\label{cc}
\end{equation}
where we dropped the third term in the left hand side since it behaves as an exact differential.

By following the standard Jacobi procedure \cite{A89} to reduce our variational principle to a geodesic one, we set ${x^a}^{\prime} \equiv g^{ab}p_b$, and by the Hamiltonian equation (\ref{s}) we obtain the metric
\begin{equation}
g^{\xi \xi} =\frac{f^{\prime}}{E}\left({\xi}^2 -1\right) \, , \qquad g^{\theta \theta} =\frac{f^{\prime}}{E} \frac{1}{{\xi}^2 -1} \, .
\label{dd}
\end{equation}
By these and by the fundamental constraint relation
 \begin{equation}
\left({\xi}^2 -1\right){p_{\xi}}^2 +\frac{{p_{\theta}}^2}{{\xi}^2 -1} =E^2 ,
\label{ee}
\end{equation}
we get 
\begin{equation}
g_{ab}{x^a}^{\prime} {x^b}^{\prime} =\frac{f^{\prime}}{E} \left\{ \left({\xi}^2 -1\right){p_{\xi}}^2 +\frac{{p_{\theta}}^2}{{\xi}^2 -1}\right\}=f^{\prime}  E .
\label{ff}
\end{equation}
By the definition $x^{a\prime}= \frac{dx^a}{ds} \frac{ds}{d\eta}\equiv u^a \frac{ds}{d\eta}$, the (\ref{ff}) rewrites
\begin{equation}
g_{ab}u^a u^b \left( \frac{ds}{d\eta} \right) ^2  = f^{\prime} E ,
\label{gg}
\end{equation}
which leads to the key relation
\begin{equation}
d\eta = \sqrt{ \frac{g_{ab}u^a u^b}{f^{\prime} E }}~ds \, .
\label{hh}
\end{equation}
Indeed the expression (\ref{hh}) together with $p_{\xi} \xi^{\prime} +p_{\theta} \theta^{\prime}=Ef^{\prime}$ allows us to put the variational principle (\ref{cc}) in the geodesic form:
\begin{equation}
\delta \int  f^{\prime} E ~d\eta  = \delta \int  \sqrt{ g_{ab}u^a u^b f^{\prime} E} ~ds= \delta \int  \sqrt{ G_{ab}u^a u^b}~ds =0
\label{ii}
\end{equation}
where the metric $G_{ab} \equiv f^{\prime} E g_{ab}$ satisfies the normalization condition $G_{ab}u^a u^b=~1$ and therefore \footnote{We take the positive root since we require that the curvilinear coordinate $s$ increases monotonically with increasing value of $f$, i.e. approaching the initial cosmological singularity.} 
\begin{equation}
\frac{ds}{d\eta}=Ef^{\prime}\Rightarrow \frac{ds}{df} =E .
\label{ll}
\end{equation}
Summarizing, in the region $\Gamma_H$ the considered dynamical problem reduces to a geodesic flow on a two dimensional Riemanniann manifold described by the line element
\begin{equation}
ds^2 =E^2 \left[ 	\frac{d{\xi }^2}{{\xi}^2 -1}+  \left(\xi^2 -1\right) d {\theta }^2 \right] .
\label{mm}
\end{equation}
The above metric (\ref{mm}) has negative curvature, since the associated curvature scalar reads $R=-\frac{2}{E^2}$;
therefore the point-universe moves over a negatively curved bidimensional space on which the potential wall (\ref{a2}) cuts the region $\Gamma_{H}$. By a way completely independent of the temporal gauge we provided a satisfactory representation of the system as isomorphic to a billiard on a Lobachevsky plane \cite{A89}.

The invariant Lyapunov exponent defined as \cite{PE77} reads 
\begin{equation}
\lambda_v =\sup \lim_{s\rightarrow \infty} \frac{\ln\left(Z^2+ \left(\frac{dZ}{ds}\right)^2\right)}{2s} =\frac{1}{E} > 0 \, ,
\label{ss}
\end{equation}
where $Z$ is the geodesic deviation field (Jacobi) orthogonal to the geodesic one.
When the point-universe bounces against the potential walls, it is reflected from a geodesic to another one thus making each of them unstable. By itself, the positive Lyapunov number (\ref{ss}) is not enough to ensure the system chaoticity, since its derivation remains valid for any Bianchi type model; the crucial point is that for the Mixmaster (type VIII and IX) the potential walls reduce the configuration space to a compact region ($\Gamma_H$), ensuring that the positivity of $\lambda_v$ implies a real chaotic behaviour, provided the factorized coordinate transformation in the configuration space
\begin{equation} \alpha = -e^{f\left(\tau\right)} a\left(\theta , \xi\right) \, , \qquad \beta_+ =e^{f\left(\tau\right)} b_+ \left(\theta , \xi\right) \, , \qquad \beta_- =e^{f\left(\tau\right)} b_- \left(\theta , \xi\right) \, ,  
\label{uu}
\end{equation}
where $f,a,b_{\pm}$ denote generic functional forms of the variables $\tau, \theta, \xi$.

\section{Statistical Mechanics Approach} 

In order to reformulate the description of the Mixmaster stochasticity
in terms of the Statistical Mechanics, we adopt in (\ref{q}) the restricted time gauge $\tau^{\prime}=1$, leading to the 
variational principle 
\begin{equation} 
\delta \int \left(   p_{\xi} \frac{d\xi }{df } +  p_{\theta} 
\frac{d\theta}{df } 
- {\cal H}_{ADM} \right) df = 0 .
\label{px} 
\end{equation} 
In spite of this restriction, for any assigned time variable $\tau$ (i.e. $\eta$) there  
exists a corresponding function $f \left(\tau \right)$ 
(i.e. a set of Misner-Chitr\'e-like variables able to 
provide the scheme presented in Section \ref{hamiltonian}) defined by the (invertible) relation 
\begin{equation} 
\frac{df }{d\tau} = \frac{{\cal H}_{ADM}}{12 D }N\left(\tau \right) e^{-2f } . 
\label{qx} 
\end{equation} 
As a consequence of the variational principle (\ref{px}) we have again the expression (\ref{u}).

In agreement with this scheme, in the region $\Gamma_H$ where the potential vanishes,
we have by (\ref{qx}) $d{\cal H}_{ADM}/df = 0$, i.e. 
$\varepsilon = E = const.$ (by other words the ADM Hamiltonian 
approaches an integral of motion). 

Hence the analysis to derive the invariant measure for the system follows 
the same lines presented in \cite{KM97,M00}. 

Indeed we got again a representation of the Mixmaster dynamics in terms of 
a two-dimensional point-universe moving within closed potential walls and over a 
negative curved surface (the Lobachevsky plane \cite{KM97}), described by the line element (\ref{mm}).
Due to the bounces against 
the potential walls and to the instability of the geodesic flow on such a plane, the dyna\-mics acquires a stochastic 
feature. This system, 
admitting an ``energy-like'' constant of motion $\left(\varepsilon = E\right)$,  
is well-described by a {\it microcanonical ensemble}, whose 
Liouville invariant measure reads 
\begin{equation} 
d\varrho = A \delta \left(E - \varepsilon \right)d\xi d\theta dp_{\xi }dp_{\theta }  \, , \qquad A=const.
\label{ux} 
\end{equation} 
where $\delta\left(x\right)$ denotes the Dirac function.
After the natural positions 
\begin{equation} 
p_{\xi } = \frac{\varepsilon}{\sqrt{\xi ^2 - 1}}\cos\phi \, , \qquad p_{\theta } = \varepsilon \sqrt{\xi ^2 - 1}\sin\phi  \, , 
\label{v} 
\end{equation} 
being $0~\le~\phi~\le~2~\pi$, 
and the integration over all 
possible values of $\varepsilon$ (depending on the initial 
conditions, they do not contain any 
information about the system chaoticity), 
we arrive to the uniform invariant measure 
\begin{equation} 
d\mu = d\xi d\theta d\phi \frac{1}{8\pi ^2} . 
\label{x} 
\end{equation} 

The validity of our potential approximation is legitimated by implementing 
the Landau-Raichoudhury theorem 
\footnote{Such a theorem, 
within the mathematical assumptions founding Einsteinian dynamics, states that 
in a synchronous reference it always exists a given instant of time in 
correspondence to which the metric determinant vanishes monotonically.} 
near the initial singularity (placed by convention in $T = 0$, 
where $T$ denotes the synchronous time, i.e. $dT=- N\left(\tau\right)d\tau$), we easily get 
that $D$ vanishes monotonically (i.e. for $T\rightarrow 0$ we 
have $d \ln D/dT > 0$). 
In terms of the adopted time variable $\tau$ 
$\left(D\rightarrow 0 \Rightarrow f(\tau) \rightarrow \infty\right)$, we have
\begin{equation} 
\frac{d \ln D}{d \tau } = 
\frac{d \ln D}{d T}\frac{dT}{d\tau } = 
- \frac{d \ln D}{d T}N\left(\tau \right)  
\label{x1} 
\end{equation}
and therefore $D$ vanishes monotonically for increasing $\tau$ as 
soon as, by (\ref{qx}), $d\Gamma / d\tau >0$. 

Now the key point of our analysis is that any stationary solution of the Liouville theorem, like (\ref{u}), remains valid 
for any choice of the time variable $\tau$; indeed in \cite{M00} the construction of the Liouville theorem with respect to the 
variables $(\xi, \theta,\phi)$ shows the existence of such properties even for the invariant measure (\ref{x}).

We conclude by remarking how, when approaching the singularity $f\rightarrow~\infty$ (i.e. ${\cal H}_{ADM}\rightarrow E$), the time gauge relation (\ref{qx}) simplifies to
\begin{equation} 
\label{qq}
\frac{df }{d\tau} = 
\frac{E e^{-2f +3\xi e^{f}}}{12 }N\left(\tau \right) e^{-2f} ; 
\label{x2} 
\end{equation} 
in agreement with the analysis presented in \cite{M00}, during a free geodesic motion the asymptotic functions 
$\xi\left(f\right), \theta\left(f\right), \phi\left(f\right)$, are provided 
by the simple system
\begin{equation}
\frac{d\xi}{df}=\sqrt{\xi^2-1}\cos\phi \, , \qquad \frac{d\theta}{df}=\frac{\sin\phi}{\sqrt{\xi^2-1}} \, ,\qquad \frac{d\phi}{df}=-\frac{\xi\sin\phi}{\sqrt{\xi^2-1}} . 
\end{equation}
Once getting $\xi\left(f\right)$ as the parametric solution
\begin{eqnarray}
\label{xx}
\xi \left(\phi\right)&=&\frac{\rho}{{\sin\phi}^2} \nonumber \\
f \left( \phi \right)&=&-a \left[ 
- \frac{1}{2} \frac{\rho \cos \phi ~{\rm arctanh} \left(\frac{1}{2} \frac{{\sin\phi}^2 +a^2 \left(1+{\cos\phi}^2 \right)}{a\rho \cos \phi}\right)}{a\rho \cos \phi} \right] +b \nonumber \\
\rho&\equiv&\sqrt{a^2 +\sin^2\phi} \, \quad a,b=const.\in \Re 
\end{eqnarray}
it reduces, for a free geodesic motion, equation (\ref{qq}) to a simple
differential one for the function $f\left(\tau\right)$. 

However, the global behaviour of $\xi$ along the whole geodesic flow, is described
by the invariant measure (\ref{x}) and therefore relation (\ref{qq}) takes a stochastic character. 
If we assign one of the two functions $f \left(\tau\right)$ or $N\left(f\right)$ 
analytically, the other one acquires a stochastic behaviour.
We see how the one-to-one correspondence between any lapse function $N\left(\eta\right)$ and the associated 
set of Misner-Chit\`e-like variables, ensures the covariant nature with respect to the time-gauge of the Mixmaster universe stochastic behaviour.\\
~\\
We are very grateful to Remo Ruffini for his valuable comments on this subject.

\end{document}